\documentclass{aa}
\usepackage[varg]{txfonts}
\usepackage{graphicx}
\usepackage{txfonts}
\usepackage{natbib,twoopt}
\bibpunct{(}{)}{;}{a}{}{,}             
\newcommand{\Mrate}{\dot{M}}
\newcommand{\Mdot}{\mbox{\,$\rm M_{\odot}$}}        
\newcommand{\Zdot}{\mbox{\,$\rm Z_{\odot}$}}        
\newcommand{\ZLMC}{\mbox{\,$\rm Z_{\rm LMC}$}}        
\newcommand{\ZSMC}{\mbox{\,$\rm Z_{\rm SMC}$}}        
\newcommand{\Ldot}{\mbox{\,$\rm L_{\odot}$}  }        
\newcommand{\aov}{$\alpha_{\rm ov}$}		
\newcommand{\amlt}{$\alpha_{\rm mlt}$}		
\newcommand{\asemi}{$\alpha_{\rm semi}$}		
\providecommand{\keywords}[1]{\textbf{\textit{Keywords--}} #1}

\begin{document}
\titlerunning{<HD limit>}
\title{A theoretical investigation of the Humphreys-Davidson limit at high and low metallicity}

\author{Erin R. Higgins\inst{\ref{inst1}}\inst{\ref{inst2}}\inst{\ref{inst3}}\and Jorick S. Vink\inst{\ref{inst1}}}

\institute{Armagh Observatory and Planetarium, College Hill, Armagh BT61 9DG, N. Ireland\label{inst1} \and Queen's University of Belfast, Belfast BT7 1NN, N. Ireland\label{inst2} \and Dublin Institute for Advanced Studies, 31 Fitzwilliam Place, Dublin, Ireland\label{inst3} 
\\email{: erin.higgins@armagh.ac.uk; jorick.vink@armagh.ac.uk}}
\date{Received 20 December 2020 / Accepted 17 February 2020}
\abstract{Current massive star evolution grids are not able to simultaneously reproduce the empirical upper luminosity limit of red supergiants, the Humphrey-Davidson (HD) limit, nor the blue-to-red (B/R) supergiant ratio at high and low metallicity. Although previous studies have shown that the treatment of convection and semiconvection play a role in the post-main sequence (MS) evolution to blue/red supergiants, a unified treatment for all metallicities has not yet been achieved.}
{In this study, we focus on developing a better understanding of what drives massive star evolution to blue and red supergiant phases, with the ultimate aim of reproducing the HD limit at varied metallicities. We discuss the consequences of classifying B and R in the B/R ratio and clarify what is required to quantify a relatable theoretical B/R ratio for comparison with observations.}
{For solar, LMC (50\% solar), and SMC (20\% solar) metallicities, we develop eight grids of MESA models for the mass range 20-60\Mdot\ to probe the effect of semiconvection and overshooting on the core helium-burning phase. We compare rotating and non-rotating models with efficient (\asemi $=$ 100) and inefficient semi-convection (\asemi $=$ 0.1), with high and low amounts of core overshooting (\aov of 0.1 or 0.5). The red and blue supergiant evolutionary phases are investigated by comparing the fraction of core He-burning lifetimes spent in each phase for a range of masses and metallicities.} 
{We find that the extension of the convective core by overshooting \aov $=$ 0.5 has an effect on the post-MS evolution which can disable semiconvection leading to more RSGs, but a lack of BSGs. We therefore implement \aov $=$ 0.1 which switches on semiconvective mixing, though for standard \asemi $=$ 1, would result in an HD limit which is higher than observed at low Z (LMC, SMC). Therefore, we need to implement very efficient semiconvection of \asemi $=$ 100 which reproduces the HD limit at log L/\Ldot $\sim$ 5.5 for the Magellanic Clouds while simultaneously reproducing the Galactic HD limit of log L/\Ldot $\sim$ 5.8 naturally. The effect of semiconvection is not active at high metallicities due to the depletion of the envelope structure by strong mass loss such that semiconvective regions could not form.}{Metallicity dependent mass loss plays an indirect, yet decisive role in setting the HD limit as a function of Z. For a combination of efficient semiconvection and low overshooting with standard $\Mrate$(Z), we find a natural HD limit at all metallicities.}

\maketitle
\section{Introduction}\label{Intro}
The maximum luminosity (L$_{\rm max}$) of red supergiants (RSGs) is an important tracer for luminous stellar populations of galaxies \citep{Massey}. This limit implies that above a certain luminosity, massive stars do not evolve to the cool supergiant phase but remain compact evolving towards a blue supergiant (BSG) or Wolf-Rayet (WR) star. \cite{HD79} showed that the maximum luminosity of RSGs, now recognised as the HD limit, was log L/\Ldot $\sim$ 5.8 for the Milky Way, \citep[see also][]{LF88, DCBsgs}. 

The HD limit actually has two key features within the Hertzsprung-Russell diagram, a cool temperature-independent section represented by L$_{\rm max}$, and a hot temperature-dependent section where the most massive stars stay hot and blue. Mass loss probably plays a direct role on the hot side of the HD limit through the relations with temperature, mass, luminosity, and proximity to the Eddington limit. Yet whether mass loss plays a role in setting the L$_{\rm max}$ remains unresolved. Though this will have important consequences for the progenitors of type II-P supernovae and whether the most massive stars produce direct collapse black holes. In this study we probe the effects of mass loss and mixing in setting L$_{\rm max}$. 

Studies have shown a metallicity dependence of radiative driven winds \cite[][]{A82, K87, Vink01}, which has been thought to influence the evolution to BSGs and RSGs \citep[e.g.][]{ChiosiMaeder, Massey, LF88}. There has often been an expectation that the HD limit shifts to higher luminosities at lower metallicity due to the physics of these metallicity-dependent winds.  Perhaps surprisingly, \cite{DCBsgs} recently showed the HD limit in the SMC to be similar to or even slightly lower than that of the LMC (log L/\Ldot $\sim$ 5.4-5.5), thereby challenging the dominance of line-driven winds in setting L$_{\rm max}$. \cite{DCBsgs} suggest that there is no evidence for metallicity-dependent winds to be the primary factor in setting the HD limit. In this study, we consider the potential indirect effect of stellar winds by probing its effect on internal mixing and its dependency in setting L$_{\rm max}$ via the length of time spent as a RSG or BSG as a function of metallicity.

\cite{LM95} considered when investigating a related issue, known as the blue-to-red supergiant (B/R) ratio, that the treatment of convection played a key role in the timescales of hot and cool supergiant phases during helium-burning (He-burning), particularly mixing processes such as convective overshooting and semiconvection. First studies by \cite{vanBergh68} showed that the B/R ratio steeply increases with increasing metallicity. \cite{LM95} scrutinised the B/R ratio in order to determine which physical processes may effect the evolution of O stars to red or blue supergiant phases. While an appreciation for the sensitivity of the B/R ratio to semiconvection and mass loss was made, a unique treatment of these processes has not been reached in reproducing the B/R ratio at varied metallicities. The B/R supergiant problem has been explored over the years with variations of its definition used interchangeably. This has caused inconsistencies within theoretical models and the observed number of supergiants. In particular, reproducing the number of BSGs and RSGs over a range of metallicities has proven unattainable, with some input parameters sufficiently reproducing B/R at high Z and others at low Z. This poses the question - is there a problem with observations, theory, or both? 

\cite{LM95} presented the problem of predicting the B/R ratio at various Z with theory, while \cite{DCBsgs} question the observed HD limit as inversely proportional to Z. In order to reconcile these issues between theory and observations, we need to better understand the mechanisms which drive the evolution of O stars to BSG an RSG phases. These studies may also have consequences for the red supergiant problem, reviewed by \cite{D17}, concerning the number of red supergiants detected by \cite{Smartt09} as progenitors of supernovae (SNe). Most massive stars above 8\Mdot\ will evolve as RSGs before exploding as SNe, with the type II-P SNe as the most common. \cite{Smartt09} studied the pre-supernovae data in order to analyse the progenitors of a range of SNe types. When comparing the observed mass distributions with theoretical predictions of RSG populations, it was found that there was a deficiency in SNe from stars with M$_{\rm init}$ > 17\Mdot\, often referred to as the 'red supergiant problem' \citep[see also][]{Kochanek, Davies19}.

In this study we compare variations of internal mixing with a focus on the convective core overshooting parameter (\aov) and semiconvection efficiency (\asemi). We explore a wide range of model configurations in order to best fit the evolution of RSGs for a range of metallicities. We provide a grid of galactic, LMC, and SMC models which explore the dependencies of each parameter, and discuss the consequences of our results in Sect. \ref{Discussion}.

\section{Method}\label{Method}
\subsection{MESA Stellar evolution models}
We avail of the one-dimensional stellar evolution code MESA (Modules for Experiments in Stellar Astrophysics) with version 8845 to calculate our models \citep{Pax11, Pax13, Pax15}, adopting physical assumptions from \cite{Higgins} unless specified otherwise. Following our investigation of the MS evolution of massive stars in \cite{Higgins}, we focus here on post-MS evolution. In particular we explore the role of semiconvective mixing in the final stages of H-burning and throughout core He-burning. As the convective core recedes during the MS, semiconvective regions form in the envelope above due to the change in the H/He abundance profile. We also investigate the effect of convective core overshooting during core H-burning as it will affect the size of the He core during later burning phases.

The Ledoux criterion\footnote{ The Ledoux criterion is denoted by $\nabla_{rad}$  $<$ $\nabla_{ad}$  $+$ $\frac{\phi}{\delta}$ $\nabla_{\mu}$ , but in chemically homogeneous layers where $\nabla_{\mu}$ $=$ 0 then the Schwarzschild criterion is effective. } is implemented for convection, while employing the mixing length theory as developed by \cite{CG68} with \amlt $=$ 1.5 \cite[e.g.][]{Jiang, Pax15}. We apply semiconvective mixing in post-MS phases with an efficiency, denoted by \cite{L83}, of \asemi\ varied here from 0.1-100. We include the effect of overshooting by extending the core by a fraction \aov of the pressure scale height H$_{\rm p}$, known as step overshooting. We vary this fraction from \aov $=$ 0.1 - 0.5, since a modest value of 0.1 is adopted by \cite{Gen12} while a moderate value of 0.335 is adopted by \cite{Bonn11}, but more recently values of up to 0.5 have been considered \citep[e.g.]{Higgins, Schoot18}, having consequences for an extended MS. We apply the 'Dutch' wind scheme as the mass loss recipe, with \cite{Vink01} active in hydrogen-rich (X$_{s}$>0.7) hot stars (T$_{\rm eff}$ > 10kK), and \cite{deJager} for cool stars (T$_{\rm eff}$ < 10kK). We apply rotation of $\Omega / \Omega_{crit}$ $=$ 0.4, as in \cite{Gen12}, and compare with non-rotating models. Rotational instabilities are employed in angular momentum transfer and mixing as described by \cite{Heger00}, though we exclude the effects of rotationally-induced mass loss as in \cite{Higgins, MullerVink}.

For selected initial chemical composition a scaled solar heavy element distribution as provided by \cite{GS98} has been adopted. We adopt a solar metallicity of Z = 0.02 (Y =0.28), with scaled-solar mass fractions applied to Z$_{\rm LMC}$ $=$ 0.0088 (Y$_{\rm LMC}$ =0.26) and Z$_{\rm SMC}$ $=$ 0.004 (Y$_{\rm SMC}$ =0.248) for the Magellanic Clouds. We calculate eight grids of models for each metallicity, rotating and non-rotating, high and low overshooting, efficient and inefficient semiconvective mixing. We determine that this grid specification is sufficient to distinguish which effects are predominant as well as determining which processes conflict or coalesce. We provide our full grid of models in Fig. \ref{grid}.

\subsection{Mixing processes} 
The effects of overshooting are relevant for the core H-burning phase though will have repercussions for later phases as an increased core size will determine the ($M_{cc}$) / ($M_{T}$) ratio. Semiconvection takes affect during the core He-burning phase and will dictate the envelope structure for the final phases. Rotational mixing has an effect on internal mixing, though does not influence our results in reproducing the HD limit. While we compare both non-rotating models and $\Omega / \Omega_{crit}$ $=$ 0.4 rotating models, we provide an illustration of both models throughout (e.g. figures \ref{Kippov}...\ref{t}) though it is important to note that the use of either grid provides qualitatively similar results and as such does not affect our conclusions.

The role of semiconvection applies to slow mixing in a region above the convective core where there is stable convection by the Ledoux criterion but unstable by the Schwarzschild criterion, \citep{L85}. The efficiency of this mixing is described by a diffusion coefficient which determines how rapidly mixing takes place. Semiconvection affects the hydrogen profile outside the He-burning core, changing the H/He abundance gradient which in turn alters the structure of the H envelope causing evolution to either red or blue supergiant phases. Semiconvection is implemented as a time-dependent diffusive process in MESA where the diffusion coefficient, D$_{sc}$, is calculated by equation \ref{eqsc}, as seen in \cite{L83}. 
\begin{center}
\begin{equation}
    D_{sc} = \alpha_{sc} \frac{K}{6C_{p}\rho} \frac{\nabla_{T} - \nabla_{ad}} {\nabla_{L} - \nabla_{T}}
    \label{eqsc}
\end{equation} 
\end{center}
where K is the radiative conductivity, C$_{p}$ is the specific heat at constant pressure and \asemi\ is a free parameter which dictates the efficiency of semiconvective mixing.

\cite{LM95} highlight that increased semiconvection leads to a higher number of BSGs whereas less semiconvection leads to more RSGs. This results from tests with the Schwarzschild criterion which increase the ratio of B/R. \cite{LM95} found that while the Schwarzschild criterion predicts the ratio for galactic metallicity as observed, and the Ledoux criterion can reproduce the ratio at SMC metallicity, none of the treatments tested were capable of simultaneously reconstructing the B/R ratio at both high and low metallicities. Perhaps the indirect effects of mass loss and mixing with metallicity can provide a better understanding of what sets the HD limit, and whether it remains a hard boundary or merely a short-lived phase. 

\subsection{Observations in the HRD}
The observed L$_{\rm max}$ is set by the most massive RSGs and in the past has been altered due to uncertainties in distances or bolometric corrections \citep[e.g.][]{DCBsgs}, but can also be determined by the timescales of the RSG phase at these highest masses. The L$_{\rm max}$ of RSGs was observed to be log L/\Ldot\ $_{\rm max}$ $\sim$ 5.4 - 5.5 for both LMC and SMC \citep{DCBsgs}. Probability distributions suggested that the SMC L$_{\rm max}$ would be slightly lower than that of the LMC, which suggests that L$_{\rm max}$ does not increase with decreasing metallicity, and in fact it may be the inverse. Luminosity distributions from this study also highlight a cross-over from RSGs to WRs in both Magellanic Clouds at log L/\Ldot $\sim$ 5.4 - 5.5, suggesting a possible shift in evolutionary channels.

Although most RSGs in \cite{DCBsgs} were observed at log L/\Ldot\ $\sim$ 5.4 - 5.5 for both LMC and SMC, there was one object in each galaxy observed to be above the HD limit. This poses a question of the HD limit being a hard border which physically should not be passed, or an observational artefact based on short timescales spent above the observed maximum luminosity, suggesting that while the most luminous RSGs are prone to small statistics, they may spend a small fraction of core He-burning as RSGs and can be observed as such, (see Fig. \ref{LMCdata}).
\begin{figure}
\centering
\includegraphics[width = 0.49\textwidth]{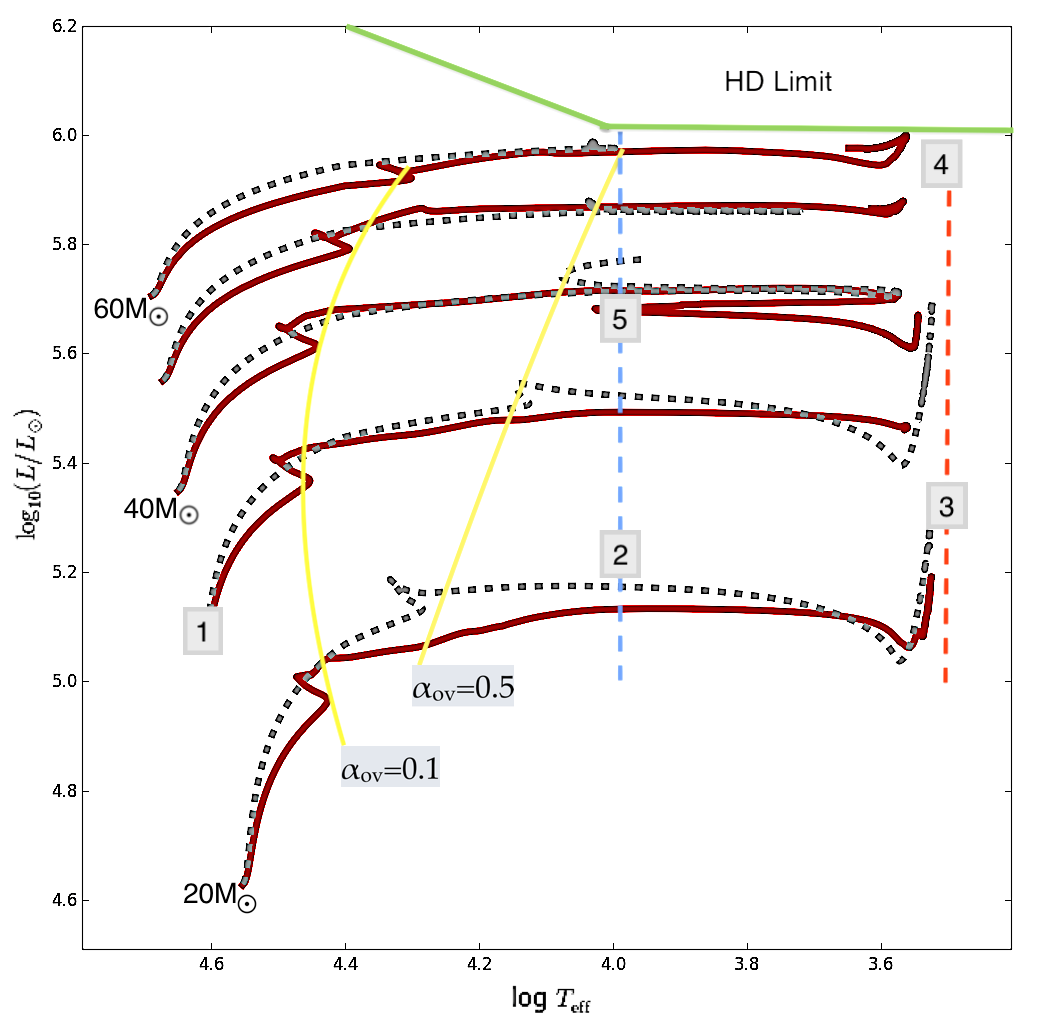}
\caption{\footnotesize Evolutionary models for 20-60\Mdot\ in HRD form, with TAMS positions shown for \aov\ $=$ 0.1, 0.5. Evolutionary phases are highlighted by number with RSGs at the red dashed line, BSGs at the blue dashed line and the MS before the yellow lines with the HD limit shown in green. Position 1 shows MS objects, positions 2 and 5 are post-MS BSGs which may be pre or post-RSGs. Position 3 represents the majority of observed RSGs, while position 4 illustrates the most luminous RSGs.}
\label{Cartoon}
\end{figure}
\section{Results}\label{Results}
We explore the effects of semiconvection and overshooting with free parameters \aov\ and \asemi\ for a range of masses 20-60\Mdot\ and metallicities \Zdot, Z$_{\rm LMC}$, and Z$_{\rm SMC}$, in order to probe the evolution to RSG or BSG phases. We assume the observed HD limit is determined by the luminosity at which massive stars spend a significant fraction of core He-burning at $\sim$ log T$_{\rm eff}$ 3.6 ($\sim$ 90$\%$ t$_{\rm He}$). We note the occurrence of blue loops and thus do not define a model as a RSG if it merely dips into the cooler temperature for a short timescale ($\sim$ 10$\%$ t$_{\rm He}$). In Sect. 3.1 we explore the evolutionary channels of massive stars which may evolve as RSGs and BSGs. We provide our results for reproducing the HD limit in sections 3.2 and 3.3, before discussing the consequences our results may have on the B/R ratio in section 3.4.

\begin{figure*}
\centering
\includegraphics[width = 18cm]{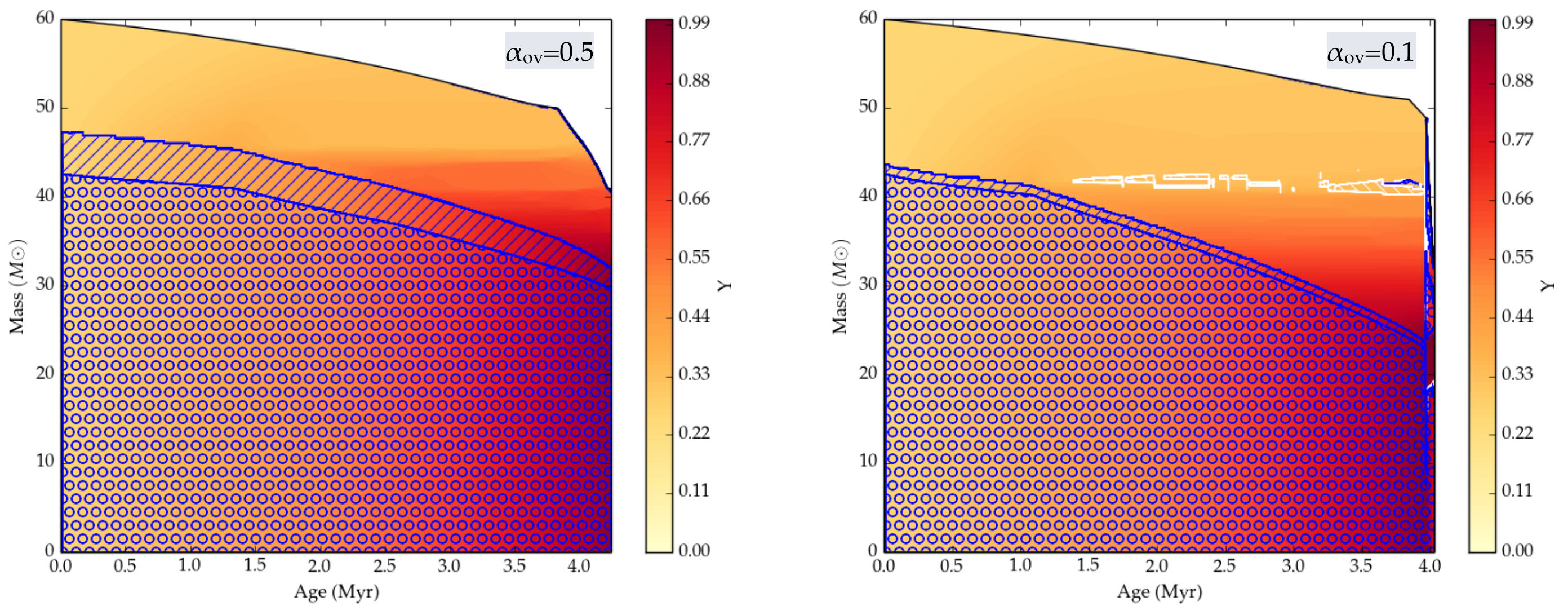}
\caption{\footnotesize Kippenhahn diagrams of 60\Mdot\ rotating models with LMC metallicity and \aov $=$ 0.1 (right) or \aov $=$ 0.5 (left). The colorbar represents the core He abundance to highlight post-MS evolution, whereas blue circles represent convective regions with hatched blue regions showing an extension of the core by overshooting. White hatched regions illustrate semiconvective mixing.}
\label{Kippov}
\end{figure*}
\subsection{Evolutionary channels}
\cite{ChiosiMaeder} characterise HRD positions by initial mass, with massive stars at M$_{\rm init}$ > 60\Mdot, moderate massive stars between 25-60\Mdot, and 'low mass' massive stars at M$_{\rm init}$ < 25\Mdot. Massive stars have significant winds which shed the outer envelope leaving an exposed core which cannot evolve to form a RSG, but more likely evolve to LBV and WR phases. Moderate massive stars form the basis of our study since these likely evolve to form RSGs since the winds are not strong enough to strip the outer envelope at early evolutionary phases. We later find that mixing and mass loss play a combined role in the duration of RSG/BSG phases (see Sect. \ref{HDsection}), though these processes can be separated in the M-L plane as described in \cite{Higgins}. Lower mass stars may evolve quickly to become a RSG during He-burning, though undergo blue loops before returning to a RSG. The extension of these blue loops is affected by mass loss and mixing. Low mass stars (M$_{\rm init}$ $\textless$ 25\Mdot) evolve to become the dominant population of RSGs which dictates the B/R ratio without reaching the HD limit. 

Figure \ref{Cartoon} illustrates that for massive stars in the mass range 20-60 \Mdot\, we observe MS objects which may include O supergiants, B supergiants and maybe even A supergiants depending on the TAMS position, (see Sect. \ref{Discussion}). The post-MS follows with higher mass ($\sim$ 40\Mdot) stars spending a large fraction of He-burning as BSGs before evolving up the Hayashi line as a RSG. These more massive RSGs determine L$_{\rm max}$. For $\sim$ 20\Mdot, the post-MS may be spent mostly as a RSG, populating the majority of RSGs in the B/R ratio.

The main sequence is represented by position 1 with ZAMS-TAMS positions in yellow, followed by BSG and RSG phases. Consequently, the B/R ratio heavily depends on the definition of a BSG and RSG. Figure \ref{Cartoon} illustrates that a BSG can be a post-MS object at either pre-RSG or post-RSG phases (positions 2 and 5 respectively). The lower mass RSGs which dominate the population are represented in position 3, while the most luminous RSGs which set the upper luminosity limit are shown in position 4. For some intermediate mass models, blue loops are observed, showing post-RSG BSGs represented by position 5. In order to distinguish which models spend a sufficient ratio of He-burning as a RSG, we compare the HRD positions from ZAMS to core He-exhaustion to observe a distinct switching of evolution from RSG to BSG and compare the time spent at hot and cool effective temperatures, see Fig. \ref{He_dotted}. 

\subsection{The HD limit}\label{HDsection}
First comparisons of theoretical models with the observed HRD by \cite{HD79} suggested an empirical boundary for the luminosity of RSGs  in the Milky Way and LMC which may be primarily due to the effects of metallicity-dependent mass loss at the highest masses. However, \cite{DCBsgs} recently found discrepancies between theory and observations at the highest luminosities suggesting that mass loss may not be primarily responsible for setting the HD limit. Population synthesis models of \cite{DCBsgs} with the GENEC code \citep{Gen12} showed a decrease in the time spent at luminosities higher than log L/\Ldot $\sim$ 5.6 during the cool supergiant phase. This suggests that the empirical boundary may be an observable artefact due to the short timescale of RSGs above L$_{\rm max}$. These models not only predict a decrease in the time spent as a RSG above a certain luminosity, but also find a higher L$_{\rm max}$ with lower metallicities due to reduced mass loss, with RSGs expected up to log L/\Ldot $\sim$ 5.7-5.8 in the SMC. Since this is not observed, theory implies that either stars do not evolve to cool supergiants at these higher luminosities, or they spend such a short time in this phase that it is not likely observed. 

After H is exhausted at the center, massive stars promptly start burning H in a shell and He in their core. This causes the radius to expand as the effective temperature cools in order to radiate the same nuclear energy from the stellar surface. Since the core mass and nuclear burning rate increases, the luminosity also increases along the Hayashi line before exploding as a SNe in the final phase of evolution. The size of the convective core during the MS ($M_{cc}$) as a fraction of the total mass ($M_{T}$) greatly dictates the post-MS evolution of massive stars since this determines how much mass remains in the envelope. Further to this, the envelope structure dictates the effectiveness of semiconvection which in turn shapes the H/He gradient. 

In \cite{Higgins} we provided an analysis of a main-sequence test-bed binary HD166734 which suggested an extension of the core by overshooting of up to \aov $=$ 0.5. For this reason, we first attempted to reproduce the HD limit with post-MS models for relatively high values of \aov $=$ 0.5. We found that when the core is extended by this fraction of the pressure scale height, the areas required to form semiconvective regions were not sufficient to reproduce any BSGs at all, even at low metallicity. Therefore only RSGs were formed, and a cut-off luminosity did not exist. In other words, we could not properly reproduce the HD limit at any metallicity. Figure \ref{Kippov} illustrates the structural changes as a result of enhanced overshooting for a 60\Mdot\ model, in which semiconvection does not occur for \aov $=$ 0.5. This is due to a lack of capacity for semiconvective regions to form leading to evolution of RSGs, even when extremely efficient semiconvection is implemented. Hence we find that increased overshooting of \aov $=$ 0.5 prevents evolution to BSGs. 

However, models with \aov $=$ 0.1 show semiconvective regions forming above the core due to the overall lowered fraction of core to total mass ratio (compared to \aov $=$ 0.5) which allows semiconvective regions to form in the envelope. These models allow for a combination of red and blue supergiants which can reproduce the HD limit, depending on the effects of their stellar winds with metallicity.

Similarly, the fraction of core mass to total mass may be altered by the natural effect that mass loss plays in depleting the envelope. In environments where the envelope loses enough mass through Z-driven winds the semiconvective regions, important for dictating RSG/BSG evolution, are prohibited from forming above the core since the envelope structure is not large enough to sustain extra mixing in these unstable regions. It is important to note that in these models, it is irrelevant whether very efficient or inefficient semiconvection is assumed since the regions are not developed \citep[c.f.]{Schoot18}. Therefore all models with moderate-high mass loss, such as \Zdot\ models, evolve to RSGs during He-burning, unless the envelope is stripped by other means causing WR or LBV-like phases. 

In order to constrain the free parameter \asemi\ and as a result obtain a better understanding of the HD limit over multiple metallicities, we explore each set of models as they evolve through core He-burning, determining the fraction of time spent at hot or cool temperatures (i.e. >10kK or <10kK) such that RSGs or BSGs would be favoured. We investigate the final status of each model (BSG/RSG) along with the time spent in each phase in order to determine whether a model would be observed as a RSG or BSG. Rather than implementing the MESA default value of unity for the semiconvection efficiency parameter \asemi, we here investigate the outcome of increasing and decreasing this by a factor of 10, similar to that of studies by \cite{Schoot19}. We find that this factor is necessary in altering the efficiency to a notable amount. 

\begin{figure*}
\centering
\includegraphics[width = 18cm]{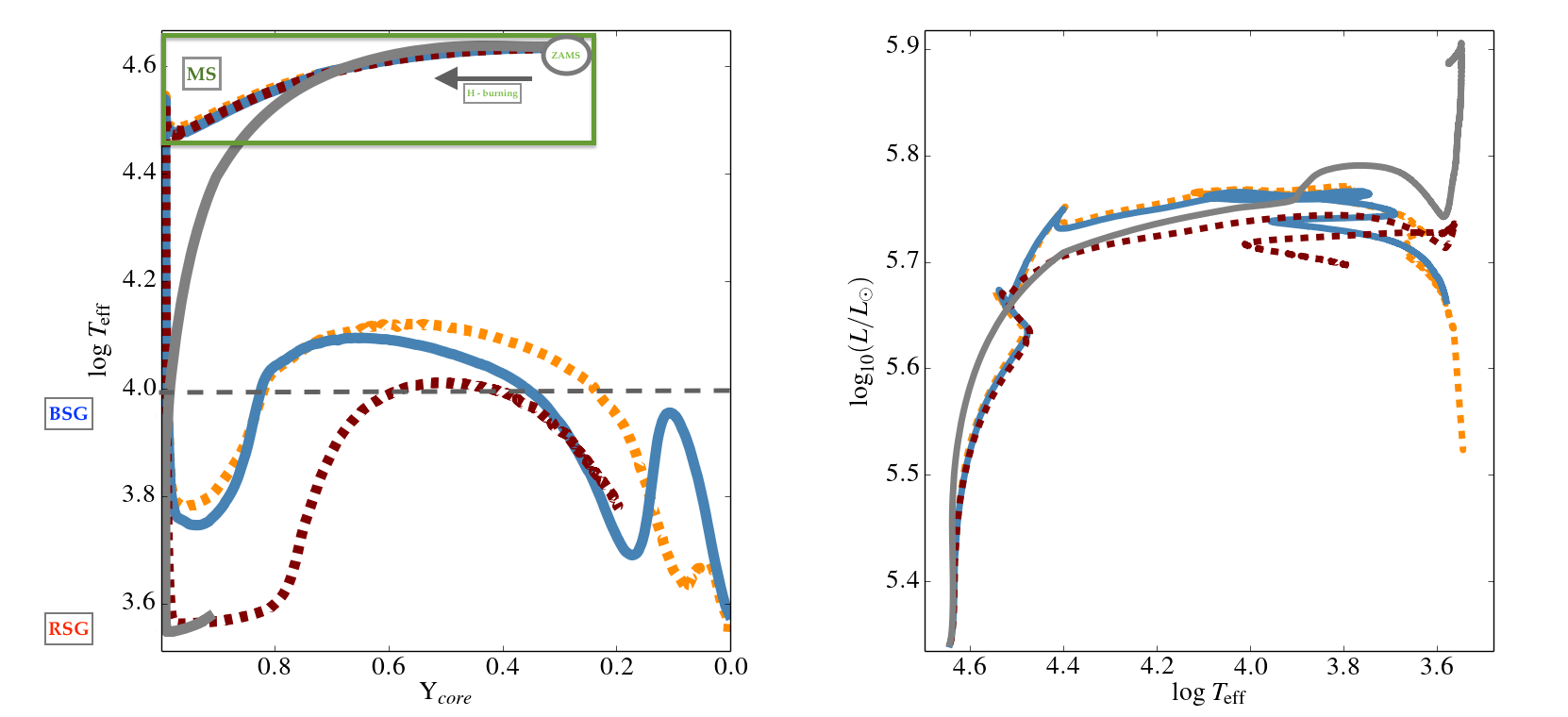}
\caption{\footnotesize Evolution of 40\Mdot\ rotating models with Z$_{LMC}$, \aov $=$ 0.1 and \asemi $=$ 1.0 (orange), 0.1 (blue), 100 (red), and finally \aov $=$0.5 with \asemi $=$ 1.0 (grey). Left: The evolution of the effective temperature as a function of central He abundance, illustrating the fraction of core He-burning spent at RSG and BSG phases. Right: Evolutionary tracks shown in a HRD highlighting a variation of blue loops and RSG evolution.}
\label{He_dotted}
\end{figure*}

An illustration of our selection criteria is shown in Fig. \ref{He_dotted}, where the HRD (right) shows whether the model evolves to the red or blue, with the core He fraction mapped out as a function of log T$_{\rm eff}$ (left). Figure \ref{He_dotted} provides a comparison of core He-burning at cool temperatures (RSG) or hot temperatures (BSG) for a 40\Mdot\ star of LMC metallicity with various factors of semiconvective efficiency. We choose this mass range as it is representative of the L$_{\rm max}$ (HD limit) where models may switch from RSG to BSG depending on input parameters. 

The main sequence is shown in the upper section of the left plot (Fig. \ref{He_dotted}) with a green box. The models then diverge at the TAMS with the grey model (\aov $=$ 0.5) leaving the main sequence later. We find that increasing the core size by overshooting \aov $=$ 0.5, which may be preferred at lower masses ($\sim$ 20-40\Mdot) as found in \cite{Higgins}, semiconvective regions are unable to form since the envelope mass is insufficient, therefore the grey model in Fig. \ref{He_dotted} evolves to a RSG and remains so until it explodes as a supernova, regardless of metallicity. 

All other models have a value of \aov $=$ 0.1 and semiconvection is varied from the default value \asemi $=$ 1 (orange). The orange dashed line remains blue through most of the He-burning phase though it will result in a RSG by core He-exhaustion. Similarly for low semiconvective efficiency, the blue line representing \asemi $=$ 0.1 spends most of core He-burning as a BSG but later dips to cooler temperatures forming a RSG. We confirm that less efficient semiconvection leads to a combination of blue and red supergiants with higher mass models reaching RSG phases but lower mass models remaining blue for most of the evolution, ending the He-burning phase as RSGs. Since observations show the opposite, i.e. more RSGs at lower masses ($\sim$ 20\Mdot) than higher masses ($\sim$ 60\Mdot), we find that this is not a solution for \asemi\ in setting the HD limit. 

For more efficient semiconvection (\asemi $=$ 100), the red dashed line shows that while this model begins He-burning as a RSG, it appears merely as a loop back into the BSG temperature range where it spends most of the He-burning timescale. Therefore when very efficient semiconvection is applied, models evolve to BSGs above a certain mass (e.g. 55\Mdot\ at \Zdot) reproducing the HD limit. 

\subsection{Unified theory of the HD limit at all Z}\label{hd}
Mass loss is decreased at lower metallicity due to its metallicity dependence, the envelope structure is large enough to create larger unstable regions which may be transformed by semiconvection. Therefore semiconvection can be more efficient in lower metallicity environments leading to more BSGs and therefore a lower HD limit. Similarly, at solar metallicity, the envelope structure would be more depleted than at LMC metallicity, prohibiting large unstable regions to develop. This means that semiconvection is overall less efficient for the same \asemi\ factor in the Galaxy than at lower metallicities. 

\begin{figure}
\centering
\includegraphics[width = 0.47\textwidth]{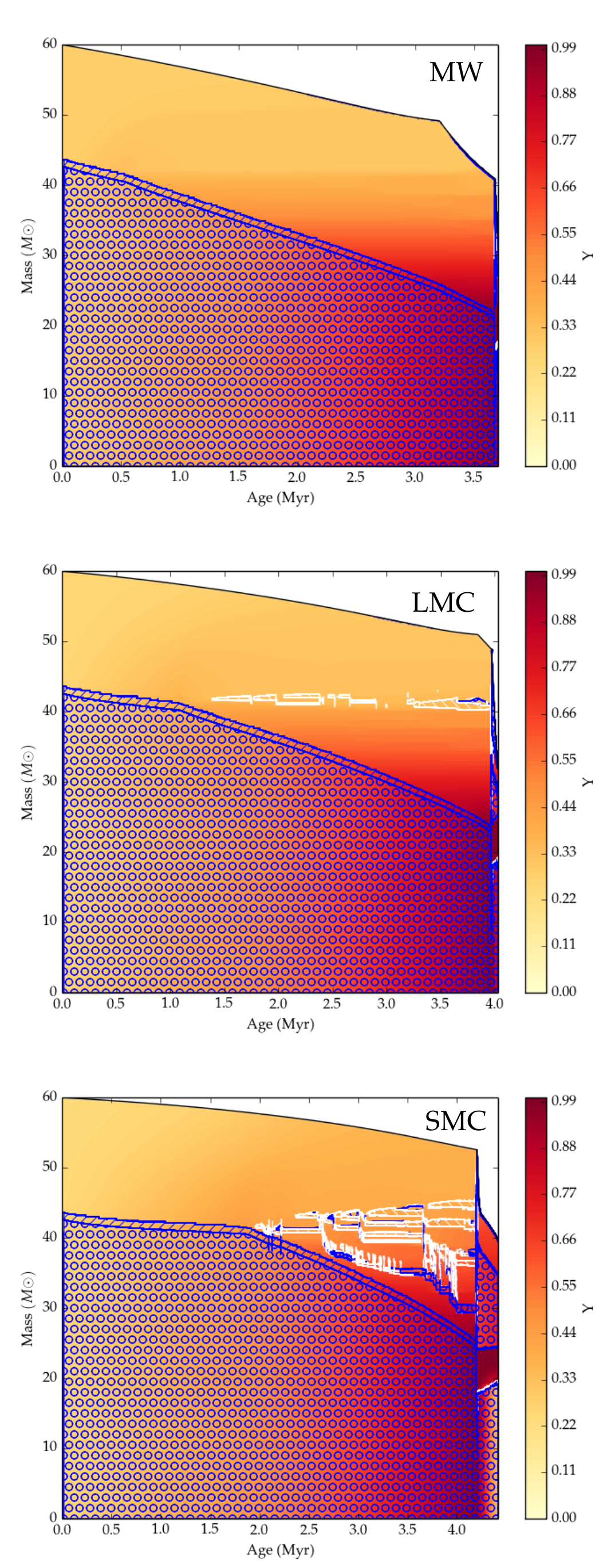}
\caption{\footnotesize Kippenhahn diagrams for 60\Mdot\ models at galactic, LMC and SMC metallicities with our preferred parameters of very efficient semiconvection and minimal overshooting (\asemi $=$ 100 and \aov $=$ 0.1, with rotation). The colorbar represents the core He abundance to highlight post-MS evolution, whereas blue circles represent convective regions with hatched blue regions showing an extension of the core by overshooting. White hatched regions illustrate semiconvective mixing. Note the decrease in envelope mass with increased metallicity, and subsequently decreased semiconvective regions.}
\label{Kipp}
\end{figure}

Figure \ref{Kipp} illustrates the extent of mass lost from the envelope of a 60\Mdot\ model with $\Delta$M $\sim$ 20\Mdot\ at solar metallicity, $\Delta$M$\sim$ 10\Mdot\ at LMC metallicity and only $\Delta$M$\sim$ 7\Mdot\ at SMC metallicity. This creates a variation in $M_{cc}$ / $M_{T}$ with metallicity leading to a higher L$_{\rm max}$ in the Milky Way than in the Magellanic Clouds. The final loss of mass during the RSG phase of evolution also has important consequences for the RSG problem and final masses which dictate the fate of these stars. 

We find that in order to reproduce the HD limit at all Z, very efficient semiconvection is needed, coupled with a small convective core or $M_{cc}$ / $M_{T}$ ratio. This can be achieved by lowering the amount of convective overshooting required at the highest masses where BSGs are expected.

Figure \ref{Kipp} demonstrates the impact of metallicity driven mass loss on semiconvection and furthermore the evolution to RSGs. Each Kippenhahn diagram in Fig. \ref{Kipp} illustrates a rotating 60\Mdot\ model with \aov $=$ 0.1 and \asemi $=$ 100 for respective metallicities. We present a 60\Mdot\ rotating model as a representation of this effect for each metallicity since mass loss plays a dominant role at this mass range. Note the increase in semiconvective regions with decreasing metallicity as a result of a larger envelope mass, for one unique set of parameters. 

These structures suggest that metallicity-dependent mass loss plays an important role in setting the HD limit as they explain the previously misunderstood increasing HD limit with increased metallicity which provided conflict between theory and observations prior to this study. We hence find an increased HD limit for increased metallicity in line with observations for our models with \asemi $=$ 100. This suggests that increased semiconvective mixing efficiency dominates the evolution, whereas mass loss plays an important indirect role in the effectiveness of semiconvection and BSG/RSG evolution. Therefore in low metallicity environments where mass loss is weaker, internal mixing is predominant.

\begin{figure}
\centering
\includegraphics[width = 0.49\textwidth]{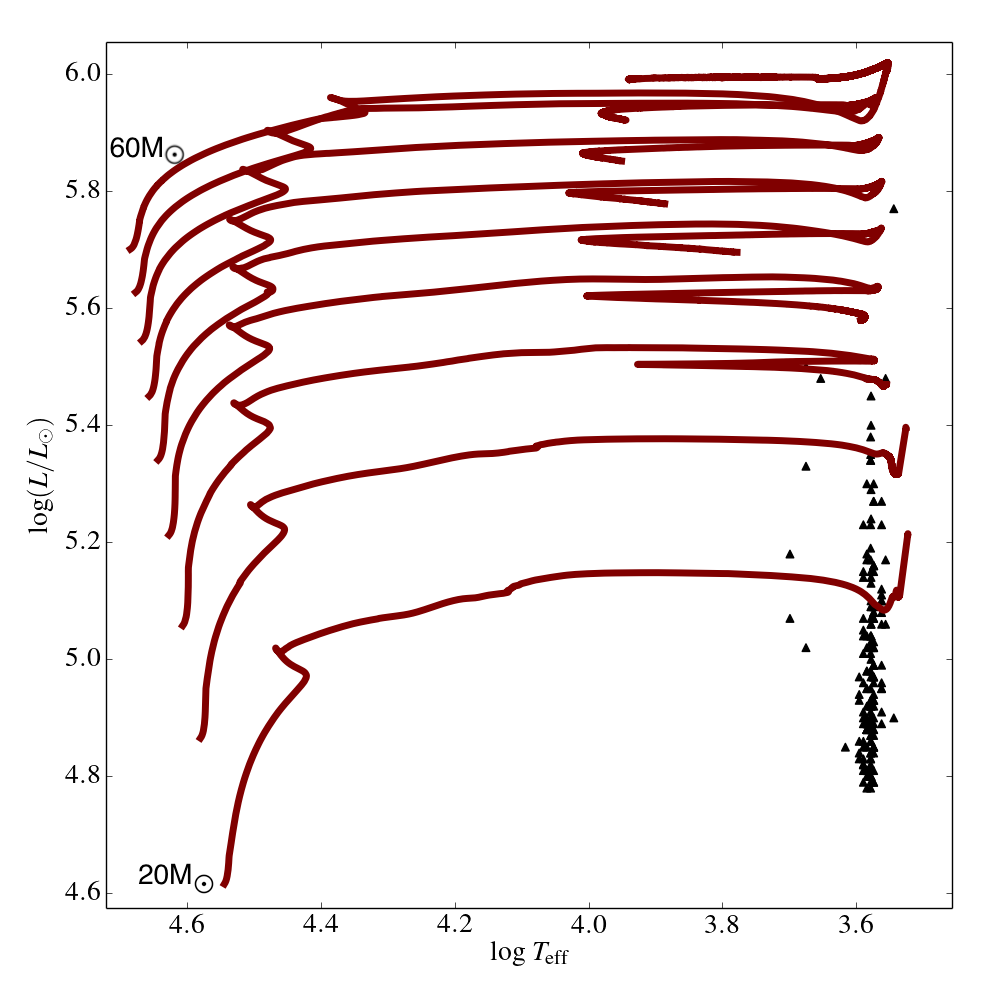}
\caption{\footnotesize Very efficient semiconvective mixing models with minimal overshooting (\aov $=$ 0.1 and \asemi $=$ 100, with rotation) for the mass range 20\Mdot\ to 60\Mdot\ (in steps of 5\Mdot) with LMC metallicity (red solid lines). The most luminous RSGs in the LMC adapted from \cite{DCBsgs} have been plotted here as black triangles as a comparison with our prescription.}
\label{LMCdata}
\end{figure}

We present the most luminous RSGs for the LMC adapted from \cite{DCBsgs} compared with our preferred set of parameters in Fig. \ref{LMCdata} with a rotating set of models of masses 20-60\Mdot\ for very efficient semiconvection. Black triangles represent the most luminous RSGs with most below log L/\Ldot $=$ 5.4. At this point our models with very efficient semiconvection begin blue loops, with masses higher than 35\Mdot\ evolving to BSGs rather than RSGs as we would expect.

Our rotating and non-rotating models with \aov $=$ 0.1 and highly efficient semiconvection (\asemi $=$ 100) are consistently able to reproduce the HD limit for three metallicities, see Fig. \ref{grid}. Rotating and non-rotating galactic metallicity models result in a HD limit of log L/\Ldot $\sim$ 5.8-5.9, in agreement with the observed maximum luminosity of RSGs in the Milky Way. Both rotating and non-rotating models of LMC and SMC metallicity find a HD limit of log L/\Ldot $\sim$ 5.4-5.5 which reproduces the distribution of RSGs in these lower metallicity environments as found by \cite{DCBsgs}. 

\begin{figure*}
\centering
\includegraphics[width = 0.9\textwidth]{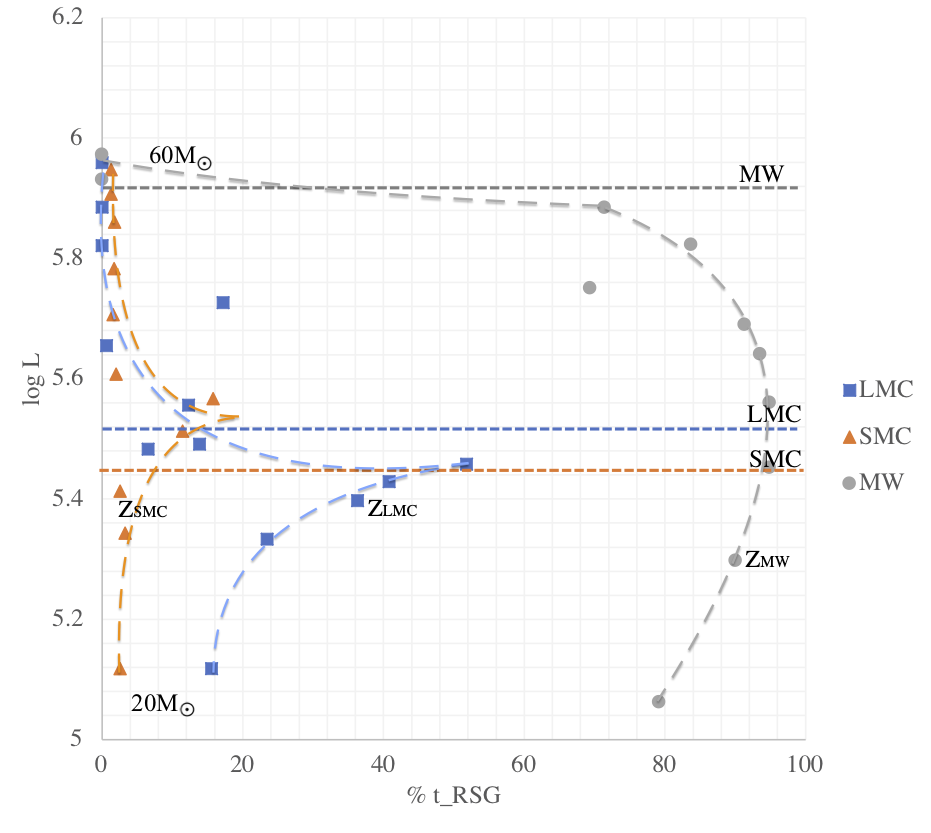}
\caption{\footnotesize The percentage of time spent during core He-burning as cool RSGs for all 20-60\Mdot\ models with \asemi$=$100 and \aov $=$ 0.1 (non-rotating). Note the dashed lines represent the observed maximum luminosity of RSGs for each galaxy / our theoretical predictions for the HD limit in each galaxy such that models which lie above these limits must spend less than $\sim$ 5\% of core He-burning at these cool temperatures.}
\label{t}
\end{figure*}

Figure \ref{t} demonstrates a theoretical HD limit based on the percentage of time spent during core He-burning at cool supergiant effective temperatures. We present our set of models from the calculated grid which could reproduce the HD limit at all metallicities. Efficient semiconvection (\asemi $=$ 100) is implemented with minimal core overshooting (\aov $=$ 0.1) for core H and He burning phases, with the indirect effects of stellar winds dominating the post-MS behaviour. Evolutionary models are analysed based on their core He-burning timescale. We compare the time spent at log T$_{\rm eff}$ > 3.6 for BSGs and < 3.6 for RSGs. Though this method determines the effective time spent as a RSG in order to determine the likelihood it would be observed above the empirical HD limit, it also provides useful information for the B/R ratio. 

We find that the galactic metallicity models spend 70-90\% of all He-burning as a RSG, leaving the L$_{\rm max}$ above log L/\Ldot\ $\sim$ 5.8. This is due the the indirect effect of stellar winds on the envelope structure at this metallicity. In the Magellanic Clouds the RSG timescales are bimodal, with lower mass RSGs (25\Mdot) spending 20-40\% of core He-burning as RSG due to their longer lifetimes, reaching 50\% at a critical point before decreasing again. The exponentially increasing and decreasing \% of time spent as a RSG form the bimodal structure seen in the LMC and SMC models. The critical point corresponds to a theoretical HD limit. The behaviour seen in the Magellanic Clouds is due to the internal mechanisms which drive evolution bluewards/redwards such as mass loss and semiconvection (see Sect. \ref{HDsection}). Above the theoretical HD limit where massive stars spend most of the He-burning phase as a BSG, the RSG timescales are <5\% of the overall He-burning time. 

For solar metallicity this only occurs at 60-70\Mdot and above log L $\sim$ 5.9, while for the Magellanic Clouds all models $\geq$ 40\Mdot\ spend less than 2\% He-burning as a RSG. Table \ref{BRtable} provides the analysis of each model shown in Fig. \ref{t}, with the timescales of core He-burning, RSG and BSG phases. Although this highlights the short timescales RSGs spend above the HD limit, it also provides the B/R ratio for a range of masses and metallicities. We find that the behaviour seen in the Magellanic clouds models, compared to the Milky Way where the behaviour is not bimodal, is a feature of semiconvective mixing being switched on and off respectively.

\begin{table*}
\caption{\label{BRtable} Timescales for core He-burning models at solar, LMC and SMC metallicity. Comparisons of time spent at hot (log T$_{\rm eff}$ $>$ 3.6) or cool (log T$_{\rm eff}$ $<$ 3.6) supergiant temperatures demonstrates the respective B/R ratio (t$_{\rm BSG}$ / t$_{\rm RSG}$) for a range of initial masses and metallicities. The luminosity is determined at the bottom of the Hayashi track.}
\centering
\begin{tabular}{c c c c c c c c}
\hline
M$_{\rm init}$ (\Mdot) &    Z & log L$_{\rm He}$ &  t$_{\rm He}$ (yrs) & t$_{\rm BSG}$ (yrs) & t$_{\rm RSG}$ (yrs) &  \% t$_{\rm RSG}$ & B/R\\
\hline
20 & \Zdot & 5.0622 & 862176 & 179694 & 682481 & 79.16 & 0.263\\
25 & \Zdot & 5.2978 & 670013 & 66537 & 6569230 & 90.07 & 0.110\\
30 & \Zdot & 5.4520 & 435955 & 22422 & 176961 & 94.86 & 0.054\\
35 & \Zdot & 5.5598 & 305292 & 15672 & 603476 & 94.87 & 0.055\\
40 & \Zdot & 5.6413 & 189269 & 12307 & 176961 & 93.50 & 0.070\\
45 & \Zdot & 5.6899 & 116566 & 10113 & 106453 & 91.32 & 0.095\\
50 & \Zdot & 5.7507 & 29658  & 9083 & 20574 & 69.37 & 0.442\\
55 & \Zdot & 5.8228 & 45261 & 7385 & 37876 & 83.68 & 0.195\\
60 & \Zdot & 5.8842 & 25656 & 7327 & 18328 & 71.44 & 0.399\\
20 & \ZLMC & 5.1178 & 917555 & 774735 & 142820 & 15.56 & 5.425\\ 
25 & \ZLMC & 5.3330 & 6956648 & 532495 & 163085 & 23.45 & 4.821\\
30 & \ZLMC & 5.4828 & 610123 & 570334 & 39789 & 6.52 & 14.334\\
35 & \ZLMC & 5.5560 & 528338 & 463547 & 64790 & 12.26 & 7.155\\
40 & \ZLMC & 5.6554 & 492070 & 488921 & 3149 & 0.64 & 155.244 \\
45 & \ZLMC & 5.7264 & 449172 & 372011 & 77160 & 17.18 & 4.821\\
50 & \ZLMC & 5.8215 & 64835 & 64834 & - & 0 & -\\
55 & \ZLMC & 5.9589 & 199940 & 199939 & - & 0 & -\\
60 & \ZLMC & 5.9800 & 57607 & 57606 & - & 0 & -\\
20 & \ZSMC & 5.1181 & 892962 & 870237 & 22724 & 2.55 & 38.29\\
25 & \ZSMC & 5.3437 & 687509 & 665022 & 22486 & 3.27 & 29.57\\
30 & \ZSMC & 5.5125 & 595742 & 527794 & 67947 & 11.41 & 7.77\\
35 & \ZSMC & 5.6076 & 524646 & 514410 & 10235 & 1.95 & 50.26\\
40 & \ZSMC & 5.7066 & 473388 & 465997 & 7390 & 1.56 & 63.05\\
45 & \ZSMC & 5.7829 & 446112 & 438579 & 7533 & 1.68 & 58.22\\
50 & \ZSMC & 5.8599 & 419623 & 4121689 & 7454 & 1.78 & 55.29\\
55 & \ZSMC & 5.9071 & 388815 & 383790 & 5025 & 1.29 & 76.38\\
60 & \ZSMC & 5.9477 & 376330 & 371621 & 4708 & 1.25 & 78.93\\
\hline
\end{tabular}
\end{table*}

\subsection{Implications for the B/R ratio}\label{B/Rsection}
In Sect. \ref{Results} we have presented our results for reproducing the HD limit at multiple metallicities while also comparing the time spent as BSGs and RSGs at the highest luminosities since this is where the HD limit is set. In the following section we discuss the full mass range (20-60\Mdot) since the RSGs formed at 20\Mdot\ dominate the B/R ratio due to the shape of the IMF.

In reproducing the HD limit from our 8 model grid sample, we also analysed the fraction of core He-burning spent as a RSG for a range of masses and metallicities, such that we could determine the likelihood that stars above the HD limit would be observed (see Fig. \ref{t}). We find that lower mass models (20...30\Mdot) have a longer He-burning timescale, spending much of the He-burning stage as a BSG before evolving redwards during the later burning phases. This results in a small percentage of He-burning as a RSG before the internal mixing affects the envelope structure dictating the time spent as a RSG or BSG. At a critical point (35-40\Mdot), models begin to spend 40-50\% of the He-burning phase as a RSG due to the effects of internal mixing, in particular efficient semiconvective mixing. The effects are predominately seen in the Magellanic clouds where mass loss is reduced. 

The behaviour changes at solar metallicity due to strong stellar winds affecting the envelope structure such that semiconvection does not take place and RSGs are formed at all masses below 60\Mdot\ (log L/\Ldot $\sim$ 5.8). The change in behaviour is present between \Zdot\ and \ZLMC\ due to the absence or presence of semiconvective regions, previously seen in Fig. \ref{Kipp}. At an intermediate metallicity this feature may switch from a high percentage of RSGs as seen at galactic metallicity to the peak seen at \ZLMC\ depending on whether semiconvective regions form or not due to the indirect effects of mass loss at a given metallicity. 

Observations from \cite{Eggen} illustrated a clear relationship between the blue-to-red supergiant ratio and metallicity for 45 open clusters, finding that the B/R ratio increased with metallicity. This study followed a series of works on the B/R ratio, though \cite{MeylanMaeder} were the first to examine the B/R ratio with stellar clusters in the Milky Way, LMC and SMC finding an increase in B/R ratio by an order of magnitude with an increase in metallicity from Z$_{\rm SMC}$ to \Zdot. Another evaluation of the B/R ratio is determined by variations in metallicity over galactocentric distance \citep[e.g.][]{vanBergh68}. This method has shown that metal-rich inner sectors of a galaxy have a higher B/R ratio than lower metallicity outer regions. \cite{HS80} demonstrated this for M33 concluding that the B/R ratio decreases with increasing galactocentric distance. It is important to note that while both sets of studies highlight a positive correlation between B/R ratio and metallicity, in this study we are concerned with defining the B/R ratio in different galaxies such as the Milky Way, LMC and SMC.

We find that evolutionary models show a decrease in time spent He-burning as a RSG with decreased Z for models 20-30\Mdot\ which corresponds to an increased B/R with decreasing Z, counter to observations when including O supergiants, (see Table \ref{BRtable}). Though as we will later discuss, the observed B/R relation with Z may be polluted by MS objects which will increase B with Z due to the varied TAMS position with Z. Results from \cite{Eggen} show that O supergiants dominate the B sample for at least seven of the 45 clusters studied, suggesting that the observed Z-relation may be a result of categorising B with MS objects as well as post-MS objects.

\section{Discussion}\label{Discussion}
The definition of B and R in an observed B/R ratio is important when comparing to theoretical models in order to establish a better understanding of the driving processes which dictate the B/R ratio such as mass loss and mixing. Since the ratio provides information on the post-MS phases we must exclude MS objects, though a lack of clarity on the TAMS position makes this exclusion difficult to determine.

The blue supergiant population is categorised by \cite{Eggen} and \cite{LM95} as O, B and A supergiants; or as O, B supergiants by \cite{MeylanMaeder}. However, O supergiants are considered H-burning objects while B and A supergiants may be either H or He-burning \citep{vink10}. By including MS objects, the overall B sample becomes much larger than R for all metallicities resulting in a high B/R ratio. 

The large number of B supergiants found adjacent to O stars in the HRD (without a gap) may suggest B supergiants are MS objects. \cite{vink10} studied the slow rotation rates of B supergiants, which could be explained by bi-stability braking in case the  wind timescale is sufficiently long, which would require that B supergiants are indeed core hydrogen-burning MS stars. This can be achieved by large core overshooting \citep{vink10}. In this case B supergiants (and perhaps even A supergiants) should not be included in the observed B/R ratio. Alternatively, the slow rotation rates of B supergiants could betray their evolved nature \citep{vink10}, in which case B sgs should be included in the B/R ratio. In other words, as long as we have not solved the related issues of the amount of core overshooting, the location of the TAMS, and the origin of the slow rotation rates of BA supergiants, we are not in a position to have a meaningful comparison of the B/R ratio between observations and theoretical model grids.

If we implement increased overshooting of \aov~$=$0.5 to allow for B supergiants as MS objects at moderate masses (20...40\Mdot) but retain a lower overshooting at higher masses (40...60\Mdot) to allow for semiconvection, then our B/R ratio can be constrained to post-MS objects which evolve to RSGs then BSGs. Due to the hotter TAMS at Z$_{\rm LMC, SMC}$ than \Zdot~ when increased overshooting is applied, the number of BSGs included as MS or post-MS changes with metallicity. This leads to more O supergiants included in B for \Zdot\ than for \ZSMC\ resulting in an increased B/R with metallicity. If we exclude MS objects from the B sample, we may exclude the Z-dependence of B/R which proves inverse to theory. This requires a detailed analysis of post-MS objects for a range of metallicities with carefully selected criteria for B and R, allowing comparisons with theoretical models. 

In Sect. \ref{Results}, we presented a unified set of input parameters which can reproduce the observed HD limit at varied metallicities. Figure \ref{grid} provides an overview these models for all three metallicities in the form of a HRD (left) and effective temperature with time (right). The core He abundance is mapped with the colourbar illustrating core He-burning at RSG and BSG effective temperatures. The model configuration included minimal core overshooting (\aov~$=$ 0.1) which would allow for semiconvective regions to form in the envelope, promoting BSG evolution. Very efficient semiconvection was also needed (\asemi~ $=$ 100) for sufficient mixing at all metallicities. We find that by core He-exhaustion, both LMC (middle) and SMC (bottom) models show a maximum luminosity of RSGs at approximately log L/\Ldot~ $\sim$ 5.5, while for solar metallicity models (top) RSGs are still formed at log L/\Ldot~ $\sim$ 5.8 during core He-burning (see Fig. \ref{grid}).

We established a second order effect in mass loss, as it plays a more significant role at \Zdot, the envelope becomes more depleted than at lower Z, resulting in fewer semiconvective regions, more RSGs and a higher observed maximum L of log L/\Ldot~ $\sim$ 5.8 (compared to that of Z$_{\rm LMC, SMC}$ where log L/\Ldot~ $\sim$ 5.5). The efficiency of semiconvection in lower z environments is higher due to the lack of mass loss even for \asemi $=$ 0.1, leading to BSGs during most of He-burning but RSGs at final stage of He-burning. 

The evolution to RSG phases has implications for pre-SNe and final mass estimates due to the strong stellar winds experienced by RSGs. The uncertainty in these mass-loss rates leads to large uncertainties in the final evolution of these stars. Figure \ref{Kippov} illustrates the final decrease in envelope mass during the RSG phase, leaving a 10\Mdot\ variation in the final mass of models with \aov $=$ 0.1 or 0.5. Since RSG mass loss rates are increased compared to earlier evolutionary phases, the consequences for increased RSG timescales can be important for determining mass-loss rates for these phases \citep[e.g.][]{Beasor18} which ultimately affects their final masses, influencing the lack of RSGs as SNe progenitors such as in the 'red supergiant problem'.

\section{Conclusions}\label{Conclusions}
We have developed eight grids of models of masses 20-60\Mdot\ for solar, LMC and SMC metallicities to probe the effect of semiconvection and overshooting on the core helium-burning phase (see Fig. \ref{grid}). We compare rotating and non-rotating models with high and low semi-convection (\asemi $=$ 0.1...100), and with high and low overshooting (\aov $=$ 0.1...0.5). We confirm that semiconvective mixing alters the envelope structure such that more blue supergiants (BSGs) are formed with more efficient semiconvection. We find that mass loss and overshooting have indirect effects which dictate the effectiveness of semiconvective regions forming, leading to more RSGs with increased mass loss and overshooting. In order to reproduce the HD limit simultaneously at all metallicities we require low \aov $\sim$ 0.1 at the mass range where the HD limit is set and above (e.g. at 55\Mdot\ and above for the Milky Way). This allows for semiconvective regions to form which in turn produce BSGs above the HD limit. 

We stress that a consistent efficiency of semiconvection may reproduce observations of the most luminous RSGs. Therefore we may constrain the HD limit solely by the efficiency of semiconvection. At higher masses we note that envelope inflation may play a role in the treatment of overshooting and the ratio of M$_{\rm cc}$ to M$_{\rm T}$, \citep[e.g.][]{GraefOwokVink}. However, since the prescription of core overshooting is uncertain we aim to better constrain the effectiveness of mixing near the core through \aov\ though it may ultimately be attributed to another internal mixing process (such as rotational mixing, perhaps mediated by internal gravity waves or dynamo mechanisms).

We appreciate that although our unique prescription of mixing and mass loss presented in this study is necessary for reproducing the HD limit at various metallicities, it will also have consequences for the B/R ratio. Since we focus on the evolution to RSG and BSG as a final stage of He-burning, the dominant mass range under scrutiny is $\sim$ 35...50\Mdot, which sets the HD limit or maximum RSG luminosity. We find that the HD limit is an observational artefact based on the likelihood of observing a RSG at such short timescales. Our models spend less than 2\%\ of core He-burning as a RSG above the theoretical HD limit for all metallicities. This suggests that while the HD limit sets a preference for BSG evolution above a certain luminosity range, it is possible to observe RSGs above the HD limit, as in \cite{DCBsgs}. 

We present an estimate of the B/R for a range of masses and metallicities, and disentangle the constraints on B and R so that observational studies may be compared to the theory which drives evolution to BSG and RSG phases. 

Our final set of models are presented in Fig. \ref{grid}, demonstrating the core He fraction timescales at RSG and BSG phases for \Zdot, \ZLMC\ and \ZSMC.

\bibliographystyle{aa} 
\bibliography{diff.bib} 

\begin{acknowledgements} 
The authors would like to thank Andreas Sander and the referee for constructive comments.
\end{acknowledgements}

\begin{appendix}\label{gridapp}\section{Grid of models}

\begin{figure}
\includegraphics[width = 0.75\textwidth]{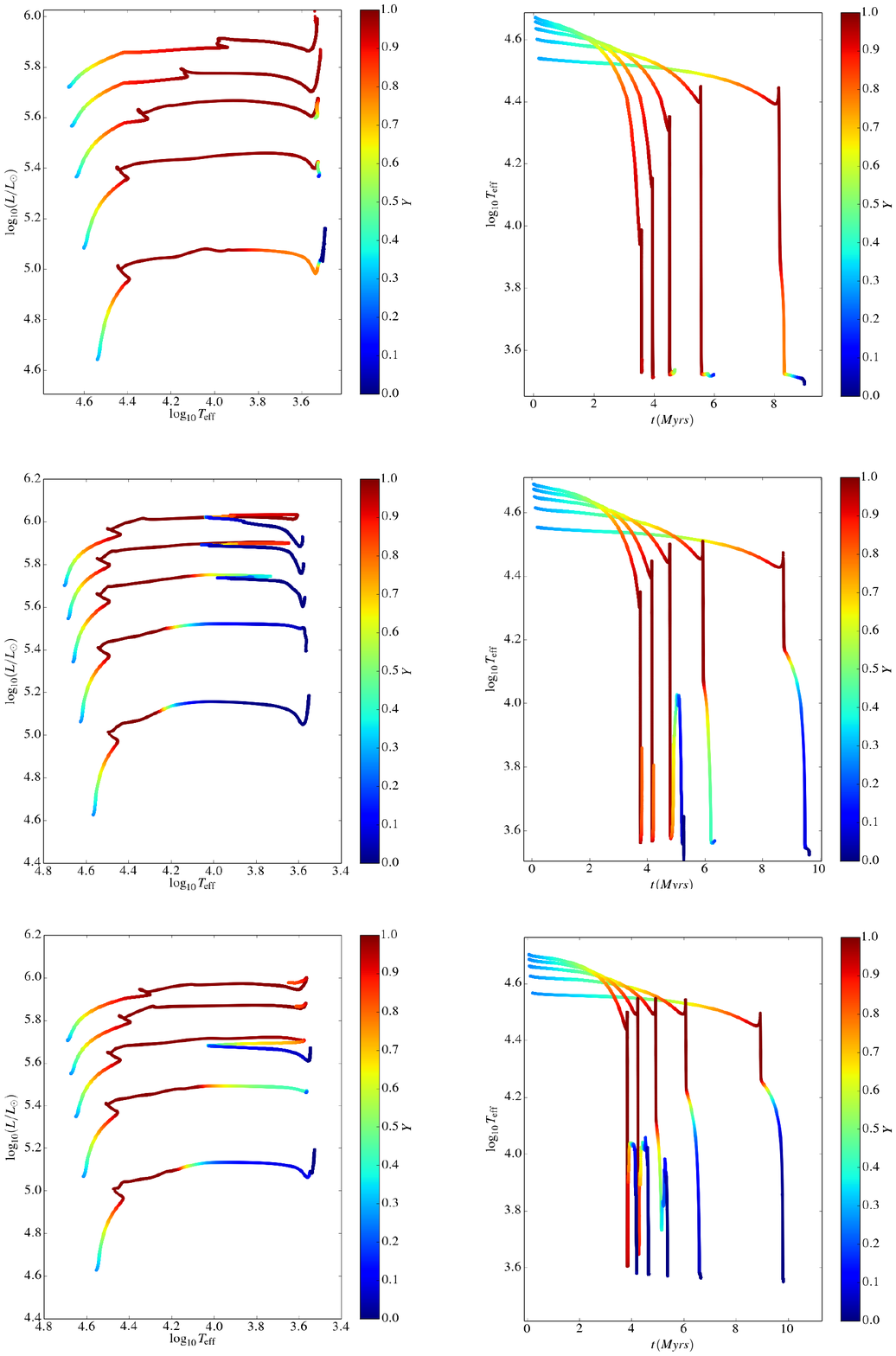}
\caption{\footnotesize Left: Evolutionary models for solar metallicity (top), LMC (middle), and SMC (bottom), with core He abundance represented by the colorbar. Right: He-burning timescale as a factor of effective temperature, such that RSGs are formed for all solar models, while only short timescales are spent in the red during He-burning at lower metallicities.}
\label{grid}
\end{figure}

\end{appendix}

\end{document}